\documentclass[
    aps,
    fleqn,
    a4paper,
    superscriptaddress,
    preprint,
    preprintnumbers,
    floatfix,
    showkeys
]{revtex4}

\usepackage{graphicx}
\usepackage{amsmath}
\usepackage[english]{babel}
\usepackage[cp1250]{inputenc}
\usepackage{epstopdf}
\usepackage{color}
\usepackage[
    colorlinks,
    linkcolor=blue,
    anchorcolor=blue,
    citecolor=blue,
    urlcolor=blue,
    menucolor=blue,
    filecolor=blue
]{hyperref}
\usepackage{subfigure}

\begin{document}

\preprint{{\em IEEE Trans. Ultrason. Ferroelectr. Freq. Control}, vol. 59, no. 9, pp. 2004, 6 Sep 2012}

\title{Application of Piezoelectric Macro-Fiber-Composite Actuators to the Suppression of Noise Transmission Through Curved Glass Plates}

\author{Kateřina \surname{Nováková}}
\email{katerina.novakova3@tul.cz} 
\affiliation{Institute of Mechatronics and Computer Engineering, Technical University of Liberec, CZ-46117 Liberec, Czech Republic}
\affiliation{Research Centre for Special Optics and Optoelectronic Systems (TOPTEC), Sobotecká 1660, CZ-51101 Turnov, Czech Republic}

\author{Pavel \surname{Mokrý}}
\affiliation{Institute of Mechatronics and Computer Engineering, Technical University of Liberec, CZ-46117 Liberec, Czech Republic}

\author{Jan \surname{Václavík}}
\affiliation{Research Centre for Special Optics and Optoelectronic Systems (TOPTEC), Sobotecká 1660, CZ-51101 Turnov, Czech Republic}

\begin{abstract}
This paper analyzes the possibility of increasing the acoustic transmission loss of sound transmitted through planar or curved glass shells using attached piezoelectric macro fiber composite (MFC) actuators shunted by active circuits with a negative capacitance. The key features that control the sound transmission through the curved glass shells are analyzed using an analytical approximative model. A detailed analysis of the particular arrangement of MFC actuators on the glass shell is performed using a finite element method (FEM) model. The FEM model takes into account the effect of a flexible frame that clamps the glass shell at its edges. A method is presented for the active control of the Young’s modulus and the bending stiffness coefficient of the composite sandwich structure that consists of a glass plate and the attached piezoelectric MFC actuator. The predictions of the acoustic transmission loss frequency dependencies obtained by the FEM model are compared with experimental data. The results indicate that it is possible to increase the acoustic transmission loss by 20 and 25~dB at the frequencies of the first and second resonant modes of the planar and curved glass shells, respectively, using the effect of the shunt circuit with a negative capacitance.
\end{abstract}

\date{6 September 2012}


\maketitle

\section{Introduction}
\label{sec:Intro}

Windows and glazed facades often represent major paths of noise transmission into a building’s interior. The noise intensity in cities is steadily increasing. Unfortunately, passive noise control methods are often not efficient, especially in the low-frequency range (up to 1~kHz). In addition, they can be difficult to apply in the case of large glazed windows or facades. At the same time, it is difficult to find a realization of active noise control that would be both efficient in a broad frequency range and financially acceptable. The control of noise transmission through large planar plates has become a challenge for scientists and technicians in the field of acoustics.

The most common methods of passive noise control are based on laminated glass technology and double glazing (see e.g., \cite{RefFahy_ZakladyIngAkustiky}). Laminated glass with different thicknesses of the interlayer and glass can improve the acoustical performance by reducing noise transmission in the fenestration system. Double-glazed windows use two separate panels of glass with an air space between them. Such a double structure has a resonance frequency which depends on the mass and distance of the plates and on the stiffness of the cavity medium. At the resonance frequency, the sound insulation is minimal. Above the resonance frequency, the soundproofing capability rises three times faster than below the resonance. Both of the aforementioned methods are efficient in reducing noise transmission at frequencies higher than 1.5 kHz. On the other hand, acoustic performance of the double-glazed window deteriorates rapidly below the double structure’s resonance, where its acoustic shielding can be even worse than that of a single glass plate. To achieve a reasonable low-frequency noise transmission suppression, heavy structures are required, leading to significant weight penalties. As a result, the search for alternative methods for low-frequency noise transmission suppression has become a very important research area.

Active noise control (ANC), in which secondary waves interfere destructively with the disturbing noise, was used for the attenuation of sound transmission through aluminum elastic plates in the works by Fuller \cite{RefFuller1} and Metcalf {\it et al.} \cite{RefFuller2}. The performance of the active system in reducing the transmitted sound was tested for several input frequencies up to 200~Hz and a reduction of about 15 to 25~dB in sound transmission was achieved. Piezoelectric actuators were investigated for use as active control inputs in \cite{RefBor-TsuenWang}. The results show that a reduction of sound transmission through the plate can be successfully achieved if the proper size, number and position of the piezoelectric actuators are chosen. A recent work by Zhu \cite{RefZhu} shows the development of thin glass panels, the vibrations of which can be controlled electronically by small rare earth voice coil actuators. The development of the control system is based on the use of a wave separation algorithm that separates the incident sound from the reflected sound. Using this method, sound transmission reduction by 10 to 15~dB in a broadband frequency range up to 600~Hz was achieved. ANC was also used to improve the sound insulation of double-glazed windows. Jakob and Möser obtained their results with a feedforward \cite{RefJakob1} and adaptive feedback controller \cite{RefJakob2}. They presented a comprehensive overview of the system with various numbers and positions of loudspeakers and microphones inside the cavity and they gave some insight into the physics of the active double-glazed window. The total sound pressure level can be reduced by nearly 8~dB and somewhat more than 5 dB with feedforward and feedback control, respectively, in the frequency range up to 400~Hz.

Later, the active structural acoustic control (ASAC ) method was developed for the reduction of the structure-borne noise. In this method, vibrating structures are used as secondary noise sources to suppress sound fields generated by primary noise sources from outside (see, e.g., Pan et al. \cite{RefA1}, Pan and Hansen \cite{RefA2}, and Hirsch et al. \cite{RefA3}). Various studies use the properties of the passive double panel system to enhance the noise control performance (e.g., Carneal and Fuller \cite{RefA4}). Naticchia and Carbonari studied the implementation of an active structural control system for glazed facades \cite{RefA5}. They obtained a sound reduction of up to 15~dB in the low-frequency range (up to 200~Hz). They have also proposed a procedure to combine this approach with laminated glass plates, which are more effective at higher frequencies. However, when applied to building windows, the ANC method requires external microphones for disturbance monitoring, and internal error sensors and loudspeakers for control purposes. To develop model-based ASAC schemes, a good understanding of structural–acoustic interactions in the considered system is required. Both of these methods require a fast control algorithm and powerful electronics. In general, such requirements yield rather expensive and energy-consuming systems.  The third category of noise suppression methods is based on the semi-active control approach. One example is an arrangement of optimally tuned Helmholtz resonators (HRs) that results in an increase in the acoustical damping level inside the cavity between the double plates. The HR is one of the most common devices for passive control of noise at low frequencies. Mao and Pietrzko \cite{RefPietrzko2005} developed a fully coupled system of structural-acoustic-HRs for double-wall structures by the modal coupling method. The simulations were confirmed by their experimental work \cite{RefPietrzko2010}. With optimally tuned HRs, it is possible to achieve sound reduction of up to 18 dB at certain low frequencies (to 100~Hz). Recent results concerning the active and passive control of sound transmission through doublewall structures have been summarized in the review by Pietrzko and Mao \cite{RefPietrzko2008}.

Another semi-active noise cancellation method with a rather high application potential is piezoelectric shunt damping (PSD). This method uses electro-acoustic transducers connected to a passive or an active shunt circuit. The approach with passive shunt circuits has been studied by Behrens et al., among others (see, e.g., \cite{RefFleming1,RefFleming2,RefFleming3}). It is mainly focused on narrow frequency band devices [based on passive resistive-inductance (RL) shunts] and relatively complicated control algorithms based on classical linear quadratic Gaussian methods (LQG) implemented through digital signal processors. Recently, reduction of noise transmission through a stiffened panel by 16~dB in the low-frequency range around the first resonance has been demonstrated by Yuan et al. \cite{RefLast1}. In this work, the results have been obtained using a hybrid control strategy combining feedforward and feedback controls. The same value of noise transmission suppression was attained by Ciminello et al. \cite{RefLast2} using a network of 6 lead zirconate titanate (PZT) actuators in multi-tone switching shunt control (SSC) system embedded in a balanced fiberglass laminate.

In this article, we have focused on active piezoelectric shunt damping (APSD), which offers an attractive approach for reduction of the noise level transmitted through windows in buildings. The APSD method is based on the change of the vibrational response of the structure using piezoelectric actuators shunted by electronic circuits which have a negative effective capacitance. It was shown by Date et al. \cite{Date2000} that the effective value of the Young’s modulus can be controlled to a large extent (from zero to infinity) by using a negative capacitor shunt. With use of this active elasticity control, resonant frequencies of the noise suppression system can be redistributed in such a way that the noise transmission through the system is greatly reduced. This approach combines the advantages of both passive and active noise control methods and it has been applied in several systems for vibration transmission suppression \cite{Kodama2002,Fukada2004apr,Ref6,Imoto2005}. Advantages of the APSD method stem from the use of inexpensive devices that are efficient in a wide frequency range (e.g., from 10~Hz to 100~kHz \cite{Tahara2006}) and have low electric energy consumption \cite{Vaclavik2012}.

The objective of this study is to analyze the most efficient ways for suppression of noise transmission through the glass plates using active elasticity control of attached piezoelectric elements. To achieve this objective, the physical quantity called acoustic transmission loss (TL) that measures the noise transmission through glass plates is introduced in Section~\ref{sec:TL}. The most important features of the noise transmission through glass plates are analyzed using the approximative analytical model presented in Section~\ref{sec:am}. Section~\ref{sec:APSD} presents the key aspects of the application of active elasticity control to noise transmission suppression through composite structures with piezoelectric layers. To verify the applicability of the APSD method to noise transmission suppression through glass plates, finite element method (FEM) simulations of sound transmission through a glass plate with attached piezoelectric elements shunted with negative capacitance circuits are performed. Section~\ref{sec:FEM} presents details of the FEM model implementation including calculation of the effective Young’s modulus of the piezoelectric macro fiber composite (MFC) actuator, calculation of the acoustic transmission loss, and the role of the boundary condition on its frequency dependence. Section~\ref{sec:setup} describes a simple experimental setup for the approximative measurements of the acoustic transmission loss. Results of the FEM model simulations and their comparison with experimental data are presented in Section~\ref{sec:results}. Finally, numerical simulations and experimental results are discussed in Section~\ref{sec:Conclusion}.

\section {Acoustic Transmission Loss of the Glass Plate}
\label{sec:TL}

\begin{figure*}[t]
\centering
	\subfigure[]{
	\includegraphics[width=60mm]{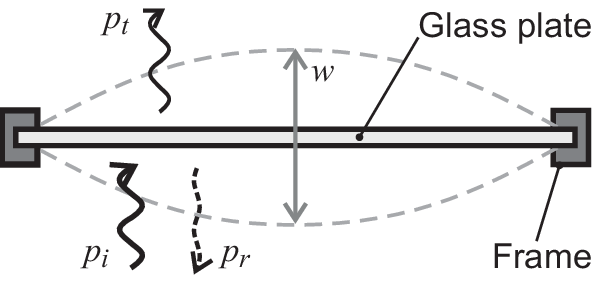}
	\label{fig:01-a}
	}
	\subfigure[]{
	\includegraphics[width=60mm]{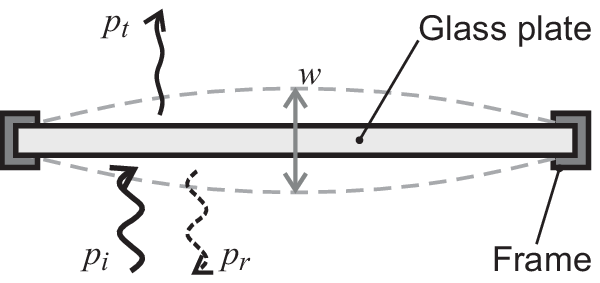}
	\label{fig:01-b}
	}
	\caption{
\subref{fig:01-a} Scheme of the considered sound transmission system, which consists of a glass plate fixed in a rigid frame at its edges. The sound source located underneath the glass plate generates an incident sound wave of acoustic pressure $p_i$ that strikes the glass plate. This makes the glass plate vibrate, and part of the sound wave is reflected and part is transmitted with the acoustic pressures $p_r$ and $p_t$, respectively. The dashed line indicates the vibration amplitude of the glass plate (for simplicity, the first mode of vibration).
\subref{fig:01-b} Scheme of the noise suppression principle: When the vibration amplitude normal to the surface of the glass plate is reduced (e.g., because it is thicker), a greater part of the incident sound wave energy is reflected than transmitted.
}
\label{fig:01}
\end{figure*}
%
The sound shielding efficiency of the window is measured by the physical quantity called acoustic transmission loss (TL). Let us consider the simple system shown in Fig.~\ref{fig:01}\subref{fig:01-a}. The system consists of a glass plate fixed in a rigid frame at its edges. The sound source located underneath the glass plate generates an incident sound wave of acoustic pressure $p_i$ that strikes the glass plate. It makes the glass plate vibrate, so that a part of the sound wave is reflected and a part is transmitted with acoustic pressures $p_r$ and $p_t$, respectively. The value of $TL$ is defined as a ratio, usually expressed in decibels, of the acoustic powers of the incident and transmitted acoustic waves, respectively.
\begin{equation}
	TL=20\log _{10}\left|\frac{p_{i}}{p_{t}}\right|.
	\label{eq:01}
\end{equation}
The value of $TL$ can be expressed in terms of the specific acoustic impedance of the window $Z_w$, as it was presented in \cite{Ref2}:
\begin{equation}
	TL=20\log _{10}\left|1+\frac{Z_{w}}{2Z_{a}}\right|,
	\label{eq:02}
\end{equation}
where $Z_a={\varrho}_0c$ is the specific acoustic impedance of air, $\varrho_0$ is the air density and $c$ is the sound velocity in the air. The specific acoustic impedance of the glass plate is defined as
\begin{equation}
	Z_w=\Delta p/v,
	\label{eq:02bZw}
\end{equation}
where $v$ and $\Delta p=(p_i+p_r)-p_t$ are the normal component of the vibration velocity and the acoustic pressure difference between the opposite sides of the glass plate, respectively.

Figure~\ref{fig:01}\subref{fig:01-b} presents a scheme of the noise suppression principle that follows from~(\ref{eq:02}) and (\ref{eq:02bZw}), i.e., the values of specific acoustic impedance $Z_w$ and acoustic transmission loss $TL$ increase with a decrease in the amplitude of the window vibration velocity $v$. To optimize the system and to achieve maximum values of $TL$, it is necessary to understand the dynamics and vibrational response of the glass plate. 

\section {Analytical Estimation of the Acoustic Impedance of a Curved Glass Shell}
\label{sec:am}

\begin{figure}[b]
\begin{center}
    \includegraphics[width=85mm]{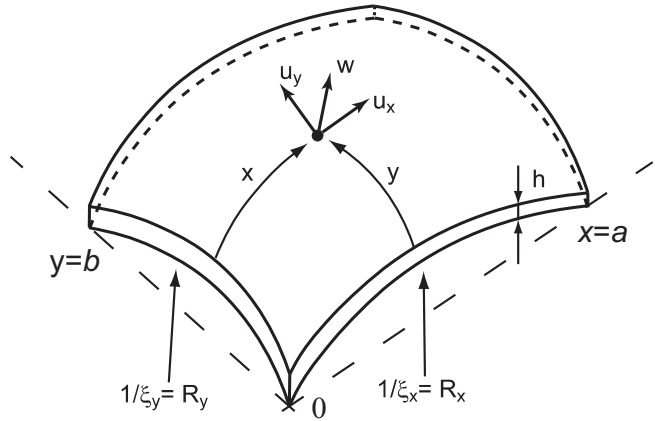}
\caption{
Geometry of the rectangular-like curved glass shell of constant thickness $h$. $x$ and $y$ are curvilinear orthogonal coordinates. The shell dimensions are denoted by $a$ and $b$. The shell has constant radii of curvatures along the $x$ and $y$ coordinates denoted by $R_x=1/\xi_x$ and $R_y=1/\xi_y$. $u_x$ and $u_y$ are tangential components of the displacement of the infinitesimal shell element. $w$ is the normal component of the displacement of the shell element.
}
\label{fig:02}
\end{center}
\end{figure}

To determine the parameters of the glass plate that control its vibrational response, the analytical model of the vibration of generally curved glass shells is developed in this section. For simplicity, consider a rectangular-like glass shell of a constant thickness $h$ and with the dimensions $a$ and $b$, as shown in Fig.~\ref{fig:02}. Consider curvilinear orthogonal coordinates $x$ and $y$ that define the position on the curved surface of the shell. Consider that the shell has constant radii of curvature along the $x$ and $y$ coordinates denoted by $R_x$ and $R_y$, respectively. It is convenient to introduce $\xi_x=1/R_x$ and $\xi_y=1/R_y$ for local curvatures of the shell along the $x$ and $y$ directions, respectively. $u_x$ and $u_y$ are tangential components of the displacement of the infinitesimal shell element with the volume $h\,dx\,dy$.  The normal component of the displacement of the shell element is denoted $w$.

%
%

Using the fundamental equations presented in basic textbooks \cite{RefB1,RefB2,RefB3}, i.e., equations expressing 1) equilibrium of forces and moments acting on an infinitesimal shell element, 2) Hooke’s law, and 3) the relations between the components of the strain tensor and the shell displacements $u_x$, $u_y$, and $w$; it is a relatively straightforward task to arrive at the following equations of motion:
\begin{subequations}
\label{eq:03}
\begin{eqnarray}
\label{eq:03a}
Yh \left[
		\frac{1}{2\left(1-\nu ^{2}\right)}
		\frac{{\partial }^{2}u_{x}}{\partial x^{2}} + 
		\frac{1}{2\left(1+\nu \right)}
		\frac{{\partial }^{2}u_{x}}{\partial y^{2}}
		- 
		\frac{1}{2\left(1-\nu \right)}
		\frac{{\partial }^{2}u_{y}}{\partial x\partial y} -
		\frac{\xi _{x}+\nu \xi _{y}}{2\left(1-\nu ^{2}\right)}
		\frac{\partial w}{\partial x}
	\right] &=& \varrho h 
	\frac{{\partial }^{2}u_{x}}{\partial t^{2}},\hspace{8mm}\\
\label{eq:03b}
Yh \left[
		\frac{1}{2\left(1+\nu \right)}
		\frac{{\partial }^{2}u_{y}}{\partial x^{2}} + 
		\frac{1}{2\left(1-\nu ^{2}\right)}
		\frac{{\partial }^{2}u_{y}}{\partial y^{2}}
		 - 
		\frac{1}{2\left(1-\nu \right)}
		\frac{{\partial }^{2}u_{x}}{\partial x\partial y} - 
		\frac{\nu \xi _{x}+\xi _{y}}{2\left(1-\nu ^{2}\right)}
		\frac{\partial w}{\partial y}
	\right] &=&
	\varrho h\frac{{\partial }^{2}u_{y}}{\partial t^{2}},\hspace{3mm}
\\
\label{eq:03c}
	-G \Delta^2w 
 + 
	\Delta p 
		\\ \nonumber 
 + 
	\frac{\mathit{Yh}}{1-\nu ^{2}}
	\left[
		\left(\xi _{x}+\nu \xi _{y}\right)
		\frac{\partial u_{x}}{\partial x} + 
		\left(\nu \xi _{x}+\xi _{y}\right)
		\frac{\partial u_{y}}{\partial y} 
		- 
		\left(\xi _{x}^{2}+\xi _{y}^{2}+2\nu \xi _{x}\xi_{y}\right)w
	\right]&=&
	\varrho h \frac{{\partial }^{2}w}{\partial t^{2}},
\end{eqnarray}
\end{subequations}
where
\begin{equation}
 \Delta^2 w = 
	\frac{\partial ^{4}w}{\partial x^{4}} + 
	2\frac{\partial ^{4}w}{\partial x^{2}\partial y^{2}} + 
	\frac{\partial ^{4}w}{\partial y^{4}}
	\label{eq:04}
\end{equation}
is the biharmonic operator, $G$ is the bending stiffness coefficient, $Y$ is the Young’s modulus, $\varrho$  is the mass density of the material, and $\nu$ is its Poisson’s ratio; $\Delta p$ is the difference between the acoustic pressures on opposite sides of the curved shell and represents the driving force of the system.

It is seen that (\ref{eq:03a}) and (\ref{eq:03b}) represent the equations of motion for the tangential components $u_x$ and $u_y$ of the shell displacement. These are coupled with the normal component of the shell displacement $w$ and the driving force $\Delta p$ via the nonzero values of curvatures $\xi_x$ and $\xi_y$. It can be shown that for relatively small numerical values of curvatures considered in this work, the values of all terms which contain the tangential components $u_x$ and $u_y$ in (\ref{eq:03c}) are much smaller than the remaining terms with the normal component $w$ of the displacement. Under this consideration, the system of equations in (\ref{eq:03}) can be further reduced to a single partial differential equation in the form
\begin{eqnarray}
	&&-G\left(
		\frac{\partial ^{4}w}{\partial x^{4}} + 
		2\frac{\partial ^{4}w}{\partial x^{2}\partial y^{2}} + 
		\frac{\partial ^{4}w}{\partial y^{4}}
	\right)
+ 
	\frac{\mathit{Yh}}{1-\nu ^{2}}
	\left(
		\xi _{x}^{2} + \xi _{y}^{2} + 2\nu \xi _{x}\xi _{y}
	\right)w + 
	\Delta p = \varrho h \frac{{\partial }^{2}w}{\partial t^{2}}.
	\label{eq:05}
\end{eqnarray}
When one considers a simple situation with the conditions 1) the shell is formed by a rectangular part of a spherical shell, i.e., $\xi_x~=~\xi_y~=~\xi$; 2) steady state, in which the shell is driven by the pure tone of angular frequency ${\omega}$, i.e., $\Delta p(t)~=~P e^{i\omega t}$ and $w(x,~y,~t)~=~w(x,~y) e^{i\omega t}$; and 3) boundary conditions of the simple supported shell, i.e., $w(0,~y) = w(a,~y) = w_{xx}(0,~y) = w_{xx}(a,~y) = 0$ and $w(x,~0) = w(x,~b) = w_{yy}(x,~0) = w_{yy}(x,~b) = 0$; the solution of the partial differential equation (\ref{eq:04}) can be easily found in the form of Fourier series, see (\ref{eq:06}), above.
%
\begin{eqnarray}
	\label{eq:06}
&&	w(x,y,t) = 
\\ \nonumber
&&	\sum _{n,m=1}^{\infty }{
		\frac{
			16P(1-\nu )
			\sin\left[(2n-1)\pi x/a\right]
			\sin\left[(2m-1)\pi y/b\right]
			e^{i\omega t}
		}{
			(2n-1)(2m-1)\pi ^{2}
			\left\{
				2 Yh \xi ^{2}+(1-\nu)
				\left[
					G\left(
						(2m-1)^{2}/b^{2}+(2n-1)^{2}/a^{2}
					\right) - 
					\rho h\omega ^{2}
				\right]
			\right\}
		}
	}.
\end{eqnarray}
%
Now, the effective value of the specific acoustic impedance $Z_w$ can be expressed in the form
\begin{equation}
	Z_w(\omega) 
	\approx 
	\Delta p(0)\left[
		i \omega 
		\sqrt{
			\frac{1}{ab} \int^{0}_{a} dx \int^{0}_{b} w(x,y,0)^2 dy
		}
	\right]^{-1}.
	\label{eq:Zw(f)_eff}
\end{equation}
%
%
\begin{eqnarray}
	Z_w(\omega) 
	\approx 
	\left\{
		\sum _{n,m=1}^{\infty } \left[
			\frac{
				8 i\omega a^2b^2(1-\nu)
			}{
				(2m-1)(2n-1)\pi^2
				\left(
					G\, \zeta_{mn} + 2Yh\xi^2 - (1-\nu)\rho h \omega^2
				\right)
			}
		\right]^2
	\right\}^{-1/2},
\label{eq:Zw(f)_eff_form}
\end{eqnarray}
%
When we substitute the expression for the normal displacement of the spherical shell $w$ from (\ref{eq:06}), one can arrive at (\ref{eq:Zw(f)_eff_form}), see above, the formula for the effective specific acoustic impedance, where
\begin{eqnarray}
	\zeta_{mn} 
	&=& \pi^4\, (1-\nu)^2 \, (1+\nu)\, 
	\left[
		(2m-1)^2/b^2
		+
		(2n-1)^2/a^2
	\right]^2.
	\nonumber
\end{eqnarray}

It is clear that (\ref{eq:06}) describes the displacement of the rectangular part of a spherical shell in a special situation without much practical interest. However, the presented analytical solution provides the possibility of tracing the key features of the system that can be used for the suppression of the noise transmission.

First, it is seen that with an increase of the glass shell curvature $\xi$, the term $2Yh{\xi}^2$ in the denominator of (\ref{eq:06}) increases. This decreases the amplitude of the shell displacement and, therefore, decreases the normal velocity of vibrations. As a result, the value of the specific acoustic impedance of the glass shell $Z_w$  increases with an increase in its curvature $\xi$, as can be seen in (\ref{eq:Zw(f)_eff_form}). The reason for this curvature effect is that the normal displacement of the curved shell is controlled by the in-plane stiffness in addition to the bending flexural rigidity. Second, the specific acoustic impedance $Z_w$ of the curved shell, i.e. $\xi>0$, increases with an increase in the Young’s modulus $Y$. Third, the value of $Z_w$ of the plane plate, i.e., $\xi=0$, increases with an increase in the bending stiffness coefficient $G$.

The concept presented in this article is based on active control of the Young’s modulus $Y$ and the bending stiffness coefficient $G$ of the glass shell by means of attached piezoelectric MFC actuator. (Basic information on MFC properties can be found in \cite{Ref9}.) The next section presents a method for the active control of the Young’s modulus $Y$ and the bending stiffness coefficient $G$ of the composite structure with piezoelectric layers.


\section {Active Control of Young’s Modulus and Bending Stiffness Coefficient of the Curved Glass Plate}
\label{sec:APSD}

\begin{figure}[t]
\centering
\includegraphics[width=60mm]{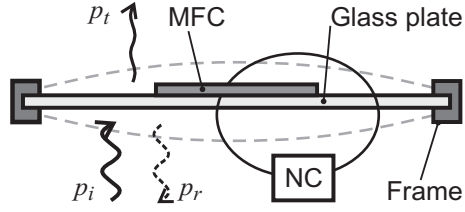}
\caption{
Scheme of the noise transmission suppression principle (cf. Fig.~\ref{fig:01}). The vibration amplitude normal to the surface of the glass plate is decreased using the effective stiffening of the piezoelectric macro fiber composite (MFC) actuator resulting from the action of the active shunt circuit with a negative capacitance (NC).
}
\label{fig:03}
\end{figure}
%
Fig.~\ref{fig:03} shows the scheme of the noise transmission suppression principle, which is achieved by using the reduction of the vibration amplitude normal to the surface of the glass plate resulting from the effective stiffening of the piezoelectric MFC actuator in the layered composite structure. The stiffening of the piezoelectric MFC actuator is realized by using an active shunt circuit with a negative capacitance (NC) according to the active elasticity control method introduced by Date et al.~\cite{Date2000}. The basic idea of the method is based on the superposition of direct and converse piezoelectric effects with Hooke’s law: First, inplane stress applied to elastic material produces strain according to Hooke’s law. In the piezoelectric MFC actuator, the in-plane stress generates a charge on its electrodes resulting from the direct piezoelectric effect. The generated charge is introduced into the shunt circuit, which controls the electric voltage applied back to electrodes of the MFC actuator. The total strain of the MFC actuator is then equal to the sum of both: the stress-induced strain (resulting from Hooke’s law) and the voltage-induced strain (resulting from the converse piezoelectric effect). When the voltage-induced strain cancels the stress-induced strain, the total strain of the MFC actuator equals zero even when nonzero stress is applied to the MFC actuator. This means that the effective Young’s modulus of the MFC actuator reaches infinity. The principle idea of the method can be simply expressed by the following formula for the effective value of Young’s modulus:
\begin{equation}
	Y_{\rm MFC} = Y_{S} \left(
		1+\frac{k^{2}}{1-k^{2}+\alpha }
	\right),
	\label{eq:07}
\end{equation}
where $Y_S$ is the Young’s modulus of piezoelectric MFC actuator material, $k$ is the electromechanical coupling factor and $\alpha=C_{\rm NC}/C_{S}$ is the ratio of the shunt circuit capacitance  $C_{\rm NC}$ over the static capacitance CS of the piezoelectric element.

\begin{figure}[t]
\centering
\includegraphics[width=80mm]{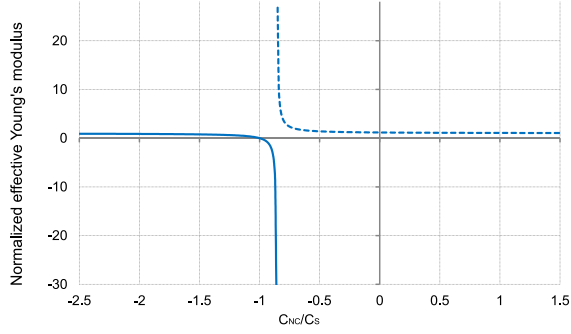}
\caption{
Normalized values of the effective Young’s modulus $Y_{\rm MFC}/Y_{S}$  of the piezoelectric MFC actuator as a function of the shunt circuit capacitance $C_{\rm NC}$ normalized to the value of the MFC actuator static capacitance $C_S$. The values of $Y_{\rm MFC}$ are calculated using (\ref{eq:07}) for $k=0.4$.
}
\label{fig:04}
\end{figure}
%
Fig.~\ref{fig:04} shows the normalized values of the effective Young’s modulus $Y_{\rm MFC}/Y_{S}$ of the piezoelectric MFC actuator as a function of the shunt circuit capacitance $C_{\rm NC}$  normalized to the value of the MFC actuator static capacitance $C_S$. The values of YMFC are calculated using (\ref{eq:07}) $k=0.4$. It follows from (\ref{eq:07}) that, when
\begin{equation}
	C_{\rm NC}=-(1-k^2)C_S, 
	\label{eq:07.5}
\end{equation}
the effective Young’s modulus reaches infinity. Such a situation has been profitably used in several noise suppression devices \cite{Okubo2001,Kodama2002,Mokry2003jul,Sluka2008}.

\begin{figure}[t]
\centering
\includegraphics[width=0.4\textwidth]{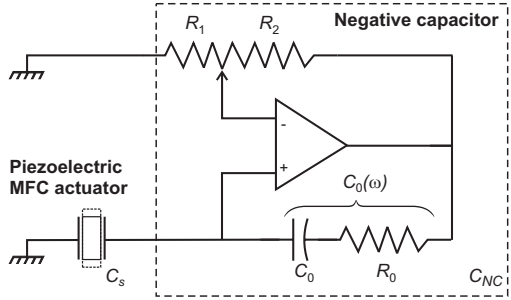}
\caption{
Electrical scheme of the piezoelectric macro fiber composite (MFC) actuator shunted by the circuit that realizes negative values of effective capacitance. $C_S$ is the piezoelectric element static capacitance, $C_{\rm 0}$ is the reference capacitance, and $R_{\rm 0}$ and $R_{\rm 1}$  are tunable resistors.
}
\label{fig:05}
\end{figure}
%
Fig.~\ref{fig:05} shows the electrical scheme of the system, in which the piezoelectric MFC actuator is shunted by the circuit that realizes negative values of capacitance. The shunt circuit with a negative capacitance is realized as a negative impedance converter circuit (i.e., a one-port circuit with an operational amplifier), where the reference impedance is realized as a capacitor $C_0$ connected in series to the resistor $R_0$. The effective value of the shunt circuit capacitance is given by
\begin{equation}
	C_{\rm NC} 
	= 
	- \left(
		\frac{
			C_0
		}{
			1 + i\omega R_0 C_0
		}
	\right)
	\frac{R_2}{R_1}.
	\label{eq:12x}
\end{equation}
By proper adjustment of the tunable resistors $R_0$ and $R_1$,  the real and imaginary part of the shunt circuit’s effective capacitance can be adjusted in such a way that the condition given by (\ref{eq:07.5}) is satisfied and the effective Young’s modulus of the MFC actuator is increased by several orders of magnitude.

\begin{figure}[b]
\centering
	\includegraphics[width=60mm]{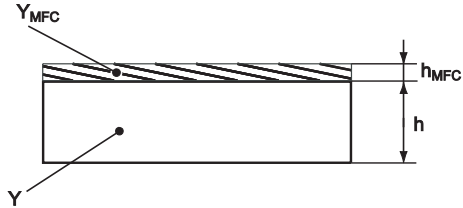}
\caption{
Cross-section of the layered composite structure, which consists of a glass plate of thickness $h$ and an attached piezoelectric macro fiber composite (MFC) actuator of thickness $h_{\rm MFC}$.
}
\label{fig:06}
\end{figure}
%
Now, let us consider the composite structure of the glass plate and the attached MFC actuator with the cross-section shown in Fig.~\ref{fig:06}. Its effective Young’s modulus $Y_{\rm Eff}$  is given by weighted average of the Young’s moduli of the glass and the MFC, according to the formula
\begin{equation}
	Y_{\rm Eff} = \frac{
		Yh+Y_{\rm MFC}h_{\rm MFC}
	}{
		h+h_{\rm MFC}
	},
	\label{eq:08}
\end{equation}
where $Y$ and $Y_{\rm MFC}$ are the Young’s moduli of the glass and the MFC actuator, respectively; $h$ and $h_{\rm MFC}$ are the thickness of the glass plate and MFC actuator, respectively.

%
\begin{equation}
	G_{\rm Eff} =
	\frac{
		Y^2 h^4 +
		Y_{\rm MFC}^2 h_{\rm MFC}^4 +
		2 Y Y_{\rm MFC} h h_{\rm MFC} 
		\left(
			2h^{2}+3hh_{\rm MFC}+2h_{\rm MFC}^{2}
		\right)
	}{
		12\left(1-\nu ^{2}\right)
		\left(
			Yh+ Y_{\rm MFC}h_{\rm MFC}
		\right)
	},
	\label{eq:09}
\end{equation}

The effective value of the bending stiffness coefficient $G$  of the composite sandwich structure is given by (\ref{eq:09}), see above, where $\nu$ is the Poisson’s ratio of the material. It is clearly seen that if Young’s modulus of the MFC is increased using the effect of the shunt circuit, both effective values of the Young’s modulus $Y_{\rm Eff}$ and the bending stiffness coefficient $G_{\rm Eff}$ of the composite sandwich structure are increased as well.

It is seen from (\ref{eq:06}) and (\ref{eq:Zw(f)_eff_form}) that with an increase of the effective Young’s modulus of the piezoelectric MFC actuator, the specific acoustic impedance of the curved glass shell increases. However, it should be noted that (\ref{eq:06}) and (\ref{eq:Zw(f)_eff_form}) were calculated in a simplified model, in which the values of curvature $\xi$, the effective Young’s modulus $Y_{\rm Eff}$ and the bending stiffness coefficient $G_{\rm Eff}$ are constant. This is not the case for many applications of practical interest. The next section presents an FEM model of a realistic system with a curved glass shell and 7 piezoelectric MFC actuators distributed on its surface. The FEM model of the system is then used for the analysis of its specific acoustic impedance.

\section {FEM Model of the Glass Plate with the Attached Shunted MFC Actuator}
\label{sec:FEM}

\begin{figure}[t]
\centering
 \includegraphics[width=0.49\textwidth]{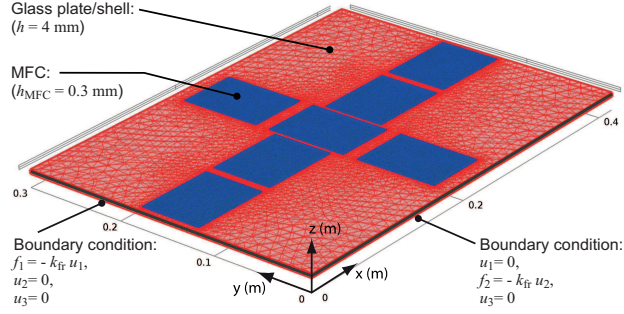}
\caption{
Geometry of the finite element method (FEM) model of a curved glass shell with 7 attached macro fiber composite (MFC) actuators. The presented configuration is selected to allow the suppression of the majority of low-frequency vibrational modes. The considered coordinate system and boundary conditions are indicated.
}
\label{fig:07}
\end{figure}
%
%
Fig.~\ref{fig:07} shows the geometry of the FEM model of a curved glass shell with 7 attached MFC actuators. We consider a glass shell which is fabricated by thermal bulging of the originally planar glass plate of thickness $h$ and dimensions $a$ and $b$. The presented configuration is selected to allow the suppression of majority of low-frequency vibrational modes. The vibrational response of the curved glass shell, expressed by the displacement vector $u_i$, is governed by the equations of motion of the form
\begin{equation}
	2 \varrho \frac{\partial ^{2}u_i}{\partial t^{2}}
	-\nabla_j\left[
		c_{ijkl} 
		\left(
			\nabla_k u_l + \nabla_l u_k
		\right)
	\right]=0,
	\label{eq:15_PDRu}
\end{equation}
where $\varrho$ is the mass density of glass, $c_{ijkl}$ are the components of the elastic stiffness tensor, and $\nabla_i=\partial/\partial x_i$  is the $i$th component of the gradient operator. Because we are interested in the steady-state vibrational response of the curved glass shell, we consider the harmonic time dependence of the displacement vector, i.e., $u_i(x,\,y,\,z,\,t)=u_i(x,\,y,\,z)\,e^{i\omega t}$, and the equations of motion (\ref{eq:15_PDRu}) can be written in the form:
\begin{equation}
	2 \omega^2 \varrho\, u_i
	+\nabla_j\left[
		c_{ijkl} 
		\left(
			\nabla_k u_l + \nabla_l u_k
		\right)
	\right]=0,
	\label{eq:16_PDRuOm}
\end{equation}

In our study, the same governing equations can be also applied to describe the vibrational response of the MFC actuators, however, with the frequency-dependent values of the components of the elastic stiffness tensor. Justification for such a simplification is as follows: Because the MFC actuator consists of the piezoelectric fibers embedded in an epoxy matrix, connected to interdigital electrodes, and laminated in thin polyimide layers, one can expect that detailed vibration of such a complicated composite structure is difficult to model in full detail. However, the MFC actuator operates as a $d_{31}$-type piezoelectric actuator with some macroscopic values of piezoelectric coefficients. In the same manner, one can introduce the effect of the shunt circuit in a manner similar to the derivation of (\ref{eq:07}) presented in \cite{Date2000}. As previously written, the effective value of Young’s modulus of the shunted MFC actuator is a function of the ratio of the shunt capacitance $C_{\rm NC}$ over the static capacitance $C_S$ of the actuator. Because the capacitance of the negative capacitor is frequency-dependent, as seen in (\ref{eq:12x}), the matching of capacitances $C_{\rm NC}$ and $C_S$ according to the condition expressed by (\ref{eq:07.5}) required for obtaining large effective values of Youngs modulus can be obtained only in a relatively narrow frequency range because of the virtually constant value of the capacitance $C_S$.

\begin{figure}[t]
\begin{center}
\includegraphics[width=85mm]{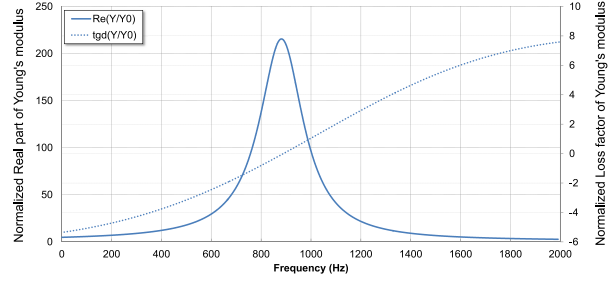}
\caption{
The effective value of the normalized real part and loss factor of the Young’s modulus of the macro fiber composite (MFC) actuator with shunted active negative capacitance (NC) circuit. Proper adjustment of the NC circuit could increase the value of the Young’s modulus by a factor of 200. The frequency 890~Hz of the peak values of the real part of the Young’s modulus is adjusted to the frequency of a particular resonant mode of the glass plate.
}
\label{fig:08}
\end{center}
\end{figure}
%
The frequency dependence of the effective Young’s modulus of the MFC actuator shunted by the negative capacitor was analyzed in detail in the recent work by Nováková and Mokrý \cite{Ref13}. Fig.~\ref{fig:08} presents the typical frequency dependence of the real part and the loss tangent of the effective value of Young’s modulus in a broad frequency range. The frequency 890~Hz of the peak values of the real part of Young’s modulus is adjusted to the frequency of a particular resonant mode of the glass plate via the particular values of the circuit parameters of the negative capacitor. In our simulations, we have therefore considered the $c_{1111}=Y_{\rm MFC}(\omega)$ and $c_{2222}=Y_{\rm MFC}(\omega)$ components of the elastic stiffness tensor of the MFC actuator as being frequency-dependent according to the function presented in Fig.~\ref{fig:08}.

Further, we consider that the glass shell with MFC actuators interacts with the acoustic field in the air above and below the shell. We consider a sound source that produces a plane incident wave below the glass shell:
\begin{equation}
	p_i(z,t)=P_i\, e^{i\left(\omega t - k z\right)},
	\label{eq:17_pi}
\end{equation}
where $k$ is the length of the wave vector of the incident sound wave, which is oriented along the z-axis. The acoustic pressure $p$ distribution in the air above and below the glass shell is governed by
\begin{equation}
	\frac{1}{\rho _{0}c^{2}}
	\frac{\partial^2 p}{\partial t^2} 
	+ 
	\nabla_i \left(-{\frac{1}{\rho _{0}}}\nabla_i p\right)
	=0,
	\label{eq:18_PRDp}
\end{equation}
where $\varrho_0$ and $c$ are the mass density and the sound speed in air. Again, we are interested in the steady-state distribution of the acoustic pressure, i.e., $p(x,\,y,\,z,\,t)=p(x,\,y,\,z)\,e^{i\omega t}$, and (\ref{eq:18_PRDp}) reduces to the form
\begin{equation}
	-\frac{\omega^2 \, p}{\rho _{0}c^{2}}
	+ 
	\nabla_i \left(-{\frac{1}{\rho _{0}}}\nabla_i p\right)
	=0.
	\label{eq:19_PRDpOm}
\end{equation}
It should be noted that below the glass shell, the acoustic pressure is given by the sum of the acoustic pressures of the incident and reflected sound waves, i.e., $p=p_i+p_r$; above the glass shell, the acoustic pressure is equal to the acoustic pressure of the transmitted sound wave, i.e., $p=p_t$. 

The calculation of the transmission loss is based on analysis of the interaction between the vibrating glass shell and the surrounding air. To facilitate the calculation, the system of partial differential equations (\ref{eq:16_PDRuOm}) and (\ref{eq:19_PRDpOm})  should be appended by the system in boundary and internal boundary conditions: First, the acoustic pressure exerts the force on the glass shell at its interface with the air, which can be expressed by the following internal boundary condition:
\begin{equation}
	n_j\left[
		c_{ijkl} 
		\left(
			\nabla_k u_l + \nabla_l u_k
		\right)
	\right]=
	2 n_i p,
	\label{eq:20_IBCpressure}
\end{equation}
where $n_i$ is the $i$th component of the outward-pointing (seen from the inside of the glass shell) unit vector normal to the surface of the glass shell with the MFC actuators. Second, the normal accelerations of the glass surface and the air particles are equal at the interfaces of the glass shell and the air:
\begin{equation}
	\omega^2\,\left(n_i u_i\right)=
	n_i\,\left(1/\varrho_0\right)\, \nabla_i p.
	\label{eq:21_IBCaccel}
\end{equation}
Finally, it will be presented later in this article that special attention must be paid to boundary conditions for the displacement $u_i$ at the edges, where the glass shell is clamped at the frame. In many real situations, the glass shell is not ideally fixed and the frame that clamps the glass shell is somewhat flexible. To take this effect into account, we approximated the boundary condition by considering a reaction force $\mathbf f_{\rm fr}$ from the spring system of the frame as being proportional to the frontal displacement of the glass shell with respect to the frame, i.e., $\mathbf f_{\rm fr} = -k_{\rm fr}\mathbf u$, where $k_{\rm fr}$ is the effective spring constant of the frame, which can be estimated from geometrical parameters and the Young’s modulus of the frame. This yields the following boundary conditions:
\begin{subequations}
\begin{eqnarray} 
	f_1 = c_{11kl} 
	\left(
		\nabla_k u_l + \nabla_l u_k
	\right)
	&=& - k_{\rm fr}\, u_1, \\
	u_2	&=& 0, \\
	u_3	&=& 0
\end{eqnarray} 
\label{eq:22_BCuAtx}
\end{subequations}
on the glass edges where $x=0$ and $x=a$ and 
\begin{subequations}
\begin{eqnarray} 
	u_1	&=& 0, \\
	f_2 = c_{22kl} 
	\left(
		\nabla_k u_l + \nabla_l u_k
	\right)
	&=& - k_{\rm fr}\, u_2, \\
	u_3	&=& 0
\end{eqnarray} 
\label{eq:23_BCuAty}
\end{subequations}
for $y=0$ and $y=b$. It should be noted that the boundary conditions given by (\ref{eq:22_BCuAtx}) and (\ref{eq:23_BCuAty}) express the limited ability of the frame to keep the glass shell edge at the same position during the vibration movements. As a result, the glass shell movements acquire some properties typical for membranes, which yield the shift of the resonant frequencies to lower values.

The boundary problem for partial differential equations given by (\ref{eq:16_PDRuOm}), (\ref{eq:19_PRDpOm})-(\ref{eq:23_BCuAty}) was solved using Comsol Multiphysics software (Comsol Inc., Burlington, MA). The solution yields spatial distributions of the acoustic pressure $p$ and the glass shell displacements $u_i$. Then, the specific acoustic impedance of the glass shell $Z_w$ was estimated for every frequency $\omega$ of the incident sound wave using the following approximative formula:
\begin{equation}
	Z_{w}(\omega)\approx\frac{\Delta{P(\omega)}}{i\omega W(\omega)},
	\label{eq:24_ZwEstim}
\end{equation}
where $\Delta P$is the amplitude of the acoustic pressure difference above and below the middle point of the glass shell; $W$ is the amplitude of the normal displacement at the middle point of the glass shell. The acoustic $TL$ was obtained using (\ref{eq:02}).

Numerical predictions of the FEM model should be compared with experimental data. The next section presents a simple setup for obtaining experimental data.

\section{Experimental Setup for the Acoustic Transmission Loss Measurements}
\label{sec:setup}

\begin{figure}[t]
\centering
\includegraphics[width=0.49\textwidth]{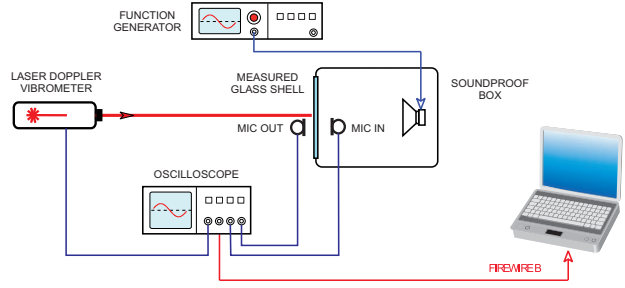}
\caption{
Experimental setup for the approximative measurements of the specific acoustic impedance. The glass plate is clamped in a wooden frame and forms a lid of the soundproof box with a loudspeaker. The microphone IN, inside the acoustic box, and the microphone OUT, outside the wooden box, measure the difference of acoustic pressures $\Delta p$ at the opposite sides of the glass plate. A laser Doppler vibrometer measures the amplitude of the vibration velocity $V$ of the middle point of the glass plate. The specific acoustic impedance $Z_w$ is approximated by the ratio $\Delta p/V$, and the value of the acoustic $TL$ is estimated using (\ref{eq:02}).
}
\label{fig:09}
\end{figure}
%
Fig.~\ref{fig:09} shows the experimental setup for the approximative measurements of the specific acoustic impedance. The glass plate is clamped in a wooden frame of the inner dimensions 0.42$\times$0.3~m. This structure forms a lid of the soundproof box with a loudspeaker that produces the source of the incident sound wave. The microphone IN, inside the box, and the microphone OUT, outside the wooden box, measure the difference of acoustic pressures $\Delta p$ at the opposite sides of the glass plate. They are placed approximately 1~cm above and below the middle point of the glass plate. A laser Doppler vibrometer measures the amplitude of the vibration velocity $V$ of the middle point of the glass plate. The specific acoustic impedance $Z_w$ is approximated by the ratio $\Delta p/V$ and the value of the acoustic $TL$ is estimated using (\ref{eq:02}).

Using such a measurement setup, only an approximative value of the $TL$ can be obtained because of the limited dimensions of the box, however, it is acceptable for the demonstration of the noise suppression efficiency. The next section presents the numerical results of our FEM model simulations and their comparison with the approximative experimental data.

\section{Results of the FEM Model Simulations and the Experimental Verification}
\label{sec:results}

\begin{figure}[b]
\begin{center}
    \includegraphics[width=0.49\textwidth]{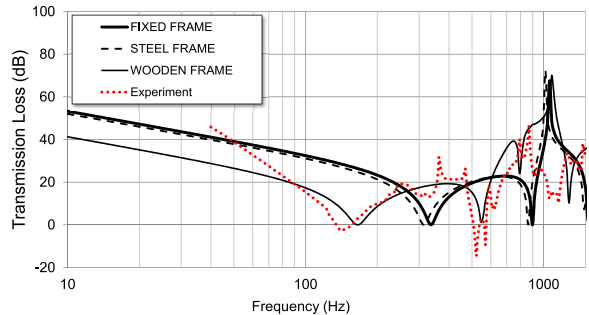}
\caption{
Comparison of the approximative measurement of the frequency dependency of (dotted line) the acoustic transmission loss through the planar glass plate with the finite element method (FEM) model predictions for three different boundary conditions: (solid thick line) ideally fixed glass, (dashed thin line) steel frame with $k_{\rm fr}=1\cdot10^{10}$~N$\cdot$m$^{-1}$, and (solid thin line) wooden frame with $k_{\rm fr}=1\cdot10^{7}$~N$\cdot$m$^{-1}$. An acceptable agreement of the experimental data with the wooden flexible frame is observed.
}   
\label{fig:10}
\end{center}
\end{figure}
%
Fig.~\ref{fig:10} shows the frequency dependencies of the acoustic TL obtained from the FEM model simulations. The following numerical parameters were considered in the numerical simulations: $P_i=0.2$~Pa, corresponding to 80~dB of the sound pressure level, $a=0.42$~m, $b=0.3$~m, and $h=4$~mm. The Young’s modulus of the glass is $73.1\cdot10^9$~Pa, the Poisson’s ratio of the glass is 0.17, and the mass density of the glass is $2203$~kg$\cdot$m$^{-3}$. The Young’s modulus of the electrically opened piezoelectric MFC actuator is equal to $Y_S=30.4\cdot 10^9$~Pa. Three situations with different boundary conditions of the glass plate were considered: (solid thick line) ideally fixed glass, (dashed thin line) steel frame with $k_{\rm fr}=1\cdot10^{10}$~N$\cdot$m$^{-1}$, and (solid thin line) wooden frame with $k_{\rm fr}=1\cdot10^{7}$~N$\cdot$m$^{-1}$. It is seen that the steel frame represents a very good approximation of the ideally fixed glass plate. On the other hand, the smaller value (by 3 orders of magnitude) of the effective spring constant of the frame causes a reasonable decrease in the resonant frequencies of the resonant modes of the glass plate. An acceptable agreement between the experimental data and the wooden flexible frame is observed. The acquired agreement of the FEM model prediction with the approximative experimental data indicates that the developed FEM model is credible and that it may provide valuable predictions of the effect of piezoelectric MFC actuators shunted by negative capacitance circuits on the frequency dependence of the acoustic transmission loss.

\begin{figure}[t]
\begin{center}
    \includegraphics[width=85mm]{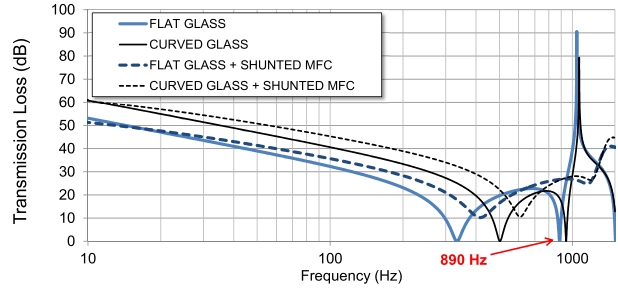}
\caption{
Frequency dependencies of the acoustic transmission loss ($TL$) obtained from the finite element method (FEM) model simulations. Parameters of the curves are the curvatures and the electrical conditions of the piezoelectric macro fiber composite (MFC) actuators: (solid thick line) planar glass plate with opened MFC actuator, (solid thin line) bulged glass shell with opened MFC actuator, (dashed thick line) planar glass plate with the MFC actuator shunted by a negative capacitance (NC) circuit, and (dashed thin line) bulged glass shell with the MFC actuator shunted by an NC circuit. The appreciable increase in the acoustic transmission loss by 10 to 25~dB in the frequency range below 400~Hz resulting from the small increase in the curvature of the glass shell is noticeable. In addition, the FEM model predictions indicate that it is possible to further improve the $TL$ by about 15~dB by the effect of the shunt NC circuit.
}
\label{fig:11}
\end{center}
\end{figure}
%
Fig.~\ref{fig:11} shows frequency dependence of the acoustic $TL$ obtained from FEM model simulations. Four situations with different curvatures and the electrical conditions of the piezoelectric MFC actuators were considered: (solid thick line) planar glass plate with opened MFC actuator, (solid thin line) bulged glass shell with opened MFC actuator, (dashed thick line) planar glass plate with the MFC actuator shunted by NC circuit, and (dashed thin line) bulged glass shell with the MFC actuator shunted by NC circuit. Fixed boundary conditions at the edges of the glass shell are considered, i.e., $u_i=0$. The bulged shape of the glass shell was approximated using the displacement function in the $z$-axis direction in the form $z_{\rm max}\sin(\pi x/a)\sin(\pi y/b)$, where $z_{\rm max}=5$~mm.

It is presented in Fig.~\ref{fig:08} that it is possible to increase the effective value of the Young’s modulus by a factor of 100 in the frequency range from 0.75 to 1~kHz and by a factor of 200 in a narrow frequency range around 0.89~kHz using the shunt circuit with a negative capacitance. Fig.~\ref{fig:11} shows that such an increase in the effective value of the Young’s modulus of the MFC actuators has an appreciable effect on the frequency dependence of the acoustic TL through the glass shell. The numerical predictions of the FEM model indicate that it is possible increase the acoustic transmission loss by 20 and 25~dB at the frequencies of the first and second resonant modes of the planar and curved glass shells, respectively.

\section{Conclusions}
\label{sec:Conclusion}

We analyzed the possibility of increasing the acoustic transmission loss of sound transmitted through planar or curved glass shells using attached piezoelectric MFC actuators shunted by active circuits with a negative capacitance.

The analytical approximative model of the sound transmission through the rectangular spherically curved laminar composite shell was developed. The analytical model was used to determine the most important features that control the sound transmission through the shell: radius of curvature, Young’s modulus, and the bending stiffness coefficient of the shell. The method for the active control of the Young’s modulus and the bending stiffness coefficient of the composite sandwich structure that consists of a glass plate and the attached piezoelectric MFC actuator was presented. Because the analytical model is applicable only to situations with little practical interest, an FEM model of a curved glass shell with attached piezoelectric MFC actuators shunted by circuits with a negative capacitance was developed. The effect of the flexible frame that clamps the glass shell edges was taken into account. The experimental setup for the approximative measurements of the specific acoustic impedance was presented.

The predictions of the acoustic transmission loss frequency dependencies obtained by the FEM model show a good agreement with the approximative measurements. It was shown that special attention must be paid to the specification of the correct boundary conditions at the edges of the glass shell. Analysis of the effect of the active elasticity control on the acoustic transmission loss of sound was performed using the FEM model. The results indicated that it is possible to increase the acoustic transmission loss by 20 and 25~dB at the frequencies of the first and second resonant modes of the planar and curved glass shells, respectively, using the effect of the shunt circuit with a negative capacitance.

The article presented a promising approach for the suppression of noise transmission through glass plates and shells using piezoelectric MFC actuators and negative capacitance circuits. The method starts from vibrational analysis focusing on the effects of the elastic properties of the composite structure with piezoelectric layers. Using active shunt circuits with a negative capacitance, the effective elastic properties of the piezoelectric layers can be controlled to a large extent. As a result, an appreciable increase in the acoustic transmission loss through the glass shell composite can be achieved. The advantages of this method stem from its generality and simplicity, offering an efficient tool for the control of the noise transmission through glass windows, especially in the low-frequency range in which passive methods are ineffective.

\section*{Acknowledgment}
\addcontentsline{toc}{section}{Acknowledgment}

This work was supported by Czech Science Foundation Project No.: GACR 101/08/1279, co-financed from the  student grant SGS 2012/7821 Interactive Mechatronics Systems Using the Cybernetics Principles, and the European Regional Development Fund and the Ministry of Education, Youth and Sports of the Czech Republic in the Project No. CZ.1.05/2.1.00/03.0079: Research Center for Special Optics and Optoelectronic Systems (TOPTEC).

\bibliographystyle{ieeetr}
\bibliography{acoustics_apsd}

\end{document}